\documentclass{elsart}
\usepackage[english]{babel}

\usepackage{epsfig}
\usepackage{graphics}
\usepackage{cite}

\usepackage{amssymb}

\begin{document}

\begin{frontmatter}

\title{Time Dependence of the Intensity of Diffracted Radiation Produced by a Relativistic Particle Passing through a Natural or Photonic Crystal}

\author{V.G. Baryshevsky, A.A. Gurinovich}

\begin{abstract}
The formulas which describe the time evolution of radiation
produced by a relativistic particle moving in a crystal are
derived. It is shown that the conditions are realizable under
which parametric (quasi-Cherenkov) radiation, transition
radiation, diffracted radiation of the oscillator, surface
quasi-Cherenkov and Smith-Purcell radiation last considerably
longer than the time $\tau_p$ of  the particle flight through the
crystal. The results of carried out experiments  demonstrate the
presence of additional radiation peak appearing after the electron
beam has left the photonic crystal.
\end{abstract}

\end{frontmatter}

\section*{Introduction}

At present, the processes of diffracted radiation of photons by
relativistic particles passing through crystals (natural or
artificial spatially periodic structures) are intensively studied
both theoretically and experimentally. Worthy of mention are such
types of diffracted radiation as parametric (quasi-Cherenkov)
radiation and diffracted radiation of a relativistic oscillator
\cite{1,2,3}. It should be noted, however, that until now,
theoretical and experimental analysis of radiation produced by a
relativistic particle passing through a crystal has focused on
spectral-angular characteristics of radiation. Nevertheless, it
was shown in \cite{nan,nan_lanl} that because of diffraction,
photons produced through radiation in crystals have group velocity
$v^{p}_{gr}$, which is appreciably smaller than the velocity $v$
of a relativistic particle.  As a result, the situation is
possible in which radiation from the crystal still continues after
the particle has passed through it \cite{nan,nan_lanl}. This
enables studying  time evolution of the process of photon
radiation  produced during the particle transmission through the
crystal (natural or photonic), or during the particle flight along
the surface of such crystals. In the present paper the formulas
are derived, which describe the time evolution of radiation
produced by a relativistic particle moving in a crystal. It is
shown that the conditions are realizable under which parametric
(quasi-Cherenkov) radiation, transition radiation, diffracted
radiation of the oscillator, surface quasi-Cherenkov and
Smith-Purcell radiation last considerably longer than the time
$\tau_p$ of  the particle flight through the crystal, i.e.,  much
longer than $\tau_p\leq 10^{-9}$ s.

\section{Spectral-angular distribution of radiation produced by a particle transmitted through a crystal}

Let us first recall the conventional consideration of the radiation process in crystals \cite{1,6}.


Both the spectral-angular density of radiation energy per unit solid angle $W_{\vec{n}\omega}$ and the differential number of emitted photons
  $dN_{\vec{n}\omega}\omega=1/\hbar\omega\cdot W_{\vec{n}\omega}$ can be easily obtained if the field $\vec{E}(\vec{r},\omega)$ produced by a particle at a large distance $\vec{r}$ from the crystal is known \cite{3}
\begin{equation}
\label{berk_2.1}
W_{\vec{n}\omega}=\frac{er^2}{4\pi^2}\overline{\left|\vec{E}(\vec{r},\omega)\right|^2},
\end{equation}
The vinculum here means averaging over all possible states of the radiating system. In order to obtain $\vec{E}(\vec{r},\omega)$, Maxwell's equation describing the interaction of particles with the medium should be solved. The transverse solution can be found with the help of Green's function of this equation, which satisfies the expression:
\begin{equation}
\label{berk_2.2}
G=G_0+G_0\frac{\omega^2}{4\pi c^2}(\hat{\varepsilon}-1)G,
\end{equation}
$G_0$ is the transverse Green's function of Maxwell's equation at $\hat{\varepsilon}=1$. It is given, for example, in \cite{66}.

Using $G$, we can find the field we are concerned with
\begin{equation}
\label{berk_2.3} E_n(\vec{r},\omega)=\int
G_{nl}(\vec{r},\vec{r}^{\prime},\omega)\frac{i\omega}{c^2}j_{0l}(\vec{r},\omega)d^3
r^{\prime},
\end{equation}
where $n, l=x, y, z$, $j_{0l}(\vec{r},\omega)$ is the Fourier
transformation of the e-th component of the current produced by a
moving beam of charged particles (in the linear field
approximation, the current is determined by the velocity and the
trajectory of a particle, which are obtained from the equation of
particle motion in the external field, by neglecting the influence
of the radiation field on the particle motion). Under the
quantum-mechanical consideration the current $j_0$ should be
considered as the current of transition of the particle-medium
system  from one state to another.

According to \cite{3,6}, Green's function is expressed at $r\rightarrow \infty$ through the solution of homogeneous Maxwell's equations $E_n^{(-)}(\vec{r},\omega)$ containing incoming spherical waves:
\begin{eqnarray}
\label{berk_2.4}
& &\lim G_{nl}(\vec{r},\vec{r}^{\prime},\omega)=\frac{e^{ikr}}{r}\sum\limits_S e^s_n E^{(-)s*}_{\vec{k}l}(\vec{r}^{\prime},\omega),\\
& &r\rightarrow \infty\nonumber
\end{eqnarray}
where $\vec{e}^s$ is the unit polarization vector, $s-1,2$, $\vec{e}^1\perp\vec{e}^2\perp\vec{k}$.

If the electromagnetic wave is incident on a crystal of finite size, then at $r\rightarrow \infty$
\[
\vec{E}_k^{(-)s}(\vec{r}, \omega)=\vec{e}^s
e^{i\vec{k}\vec{r}}+\mbox{const}\frac{e^{ikr}}{r},
\]
and one can show that the relation between the solution
$\vec{E}_k^{(-)s}$ and the solution of Maxwell's equation
$\vec{E}^{(+)}(\vec{k}, \omega)$ describing scattering of  a plane
wave by the target (crystal), is given by:
\begin{equation}
\label{berk_2.5}
\vec{E}^{(-)s*}_{\vec{k}}=\vec{E}^{(+)s}_{-\vec{k}}
\end{equation}

Using (\ref{berk_2.3}), we obtain

\begin{equation}
\label{berk_2.6} E_n(\vec{r},
\omega)=\frac{e^{ikr}}{r}\frac{i\omega}{c^2} \sum\limits_S
e^s_n\int E^{(-)s*}_{\vec{k}}(\vec{r},
\omega)\vec{j}_0(\vec{r}^{\prime}, \omega)d^3 r^{\prime}.
\end{equation}
As a result, the spectral energy density of photons with polarization $s$ can be written in the form:
\begin{equation}
\label{berk_2.7}
W_{\vec{n},\omega}^s=\frac{\omega^2}{4\pi^2c^2}\overline{\left|\int\vec{E}^{(-)s*}_{\vec{k}}(\vec{r},
\omega) \vec{j}_0(\vec{r}, \omega)d^3 r\right|^2},
\end{equation}
\begin{equation}
\label{berk_2.8}
\vec{j}_0(\vec{r}, \omega)=\int e^{i\omega t}\vec{j}_0(\vec{r}, \omega) dt=eQ \int e^{i\omega t}\vec{v}(t)\delta(\vec{r}-\vec{r}(t))dt,
\end{equation}
where $eQ$ is the charge of the particle, $\vec{v}(t)$ and $\vec{r}(t)$ are the velocity and the trajectory of the particle at moment $t$. By introducing (\ref{berk_2.8}) into (\ref{berk_2.7}) we get

\begin{equation}
\label{berk_2.9}
dN^s_{\vec{n}, \omega}=\frac{e^2 Q^2\omega}{4\pi^2 \hbar c^3}\overline{\left|\int \vec{E}^{(-)s*}_{\vec{k}}(\vec{r}(t), \omega)\vec{v}(t) e^{i\omega t}d\right|^2} t.
\end{equation}
Integration in (\ref{berk_2.9}) is carried out over the whole interval of the particle motion. It should be noted that the application of the solution of a homogeneous Maxwell's equation  instead of the inhomogeneous one essentially simplifies the analysis of the radiation problem and enables one to consider various cases of radiation emission taking into account multiple scattering.



Using equations (\ref{berk_2.7})--(\ref{berk_2.9}), one can easily obtain the explicit expression for the radiation intensity and that for the effect of multiple scattering on the process under study \cite{3,6,7}.

Consider, for example, the PXR radiation. Let a particle moving
with a uniform velocity be incident on a crystal plate with the
thickness $L$ being $L\ll L_c$, where $L_c=(\omega q)^{-1/2}$ is
the coherent length of bremsstrahlung $q=\overline{\theta}^2/4$
and $\overline{\theta}^2$ is the mean square angle of multiple
scattering. The latter requirement allows neglecting the multiple
scattering of particles by atoms. A theoretical method describing
multiple scattering effect on the {radiation process} is given in
\cite{para_4}.

According to (\ref{berk_2.9}), in order to determine the number of
quanta emitted by a particle passing through the crystal plate,
one should first find the explicit expressions for the solutions
$\vec{E}^{(-)s}_{\vec{k}}$. As was mentioned above, the field
$\vec{E}^{(-)s}_{\vec{k}}$ can be found from the relation
$\vec{E}^{(-)s}_{\vec{k}}=(\vec{E}^{(+)s}_{-\vec{k}})^*$ if one
knows the solution $\vec{E}^{(+)s}_{\vec{k}}$ describing the
photon scattering by the crystal.

In the case of two strong waves excited under diffraction (the
so-called two-beam diffraction case \cite{132}), one can
obtain the following set of equations for determining the wave
amplitudes (see \cite{lanl_7a}):
\begin{eqnarray}
\label{para_1.15}
\left(\frac{k^2}{\omega^2}-1-\chi^*_0\right)\vec{E}^{(-)s}_{\vec{k}}c_s\chi^*_{-\vec{\tau}}\vec{E}^{(-)s}_{\vec{k}_{\tau}}=0\nonumber\\
\left(\frac{k^2}{\omega^2}-1-\chi^*_0\right)\vec{E}^{(-)s}_{\vec{k}_{\tau}}c_s\chi^*_{\vec{\tau}}\vec{E}^{(-)s}_{\vec{k}}=0.
\end{eqnarray}
Here $\vec{k}_{\vec{\tau}}=\vec{k}+\vec{\tau}$, $\vec{\tau}$ is
the reciprocal lattice vector, $\chi_0$, $\chi_{\vec{\tau}}$ are
the Fourier components of the crystal susceptibility. It is well
known that the crystal is described by a periodic susceptibility
(see, for example, \cite{132}:
\begin{equation}
\label{para_1.16}
\chi(\vec{r})=\sum_{\vec{\tau}}\chi_{\vec{\tau}}\exp(i\vec{\tau}\vec{r}).
\end{equation}
$c_s=\vec{e}^s\vec{e}^s_{\vec{\tau}}$, where
$\vec{e}^s(\vec{e}^s_{\vec{\tau}})$ are the unit polarization
vectors of the incident and diffracted waves, respectively.

The condition for the linear system (\ref{para_1.15}) to be
solvable leads to a dispersion equation that determines the
possible wave vectors $\vec{k}$ in a crystal. These wave vectors
are convenient to present in the form:
\[
\vec{k}_{\mu s}=\vec{k}+\vec{\kappa}^*_{\mu s}\vec{N},\qquad
\kappa_{\mu s}^*=\frac{\omega}{c\gamma_0}\varepsilon^*_{\mu s},
\]
where $\mu=1,2$; $\vec{N}$ is the unit vector of a normal to the
entrance crystal surface which is directed into the crystal,
\begin{eqnarray}
\label{para_1.17} &
&\varepsilon_{1(2)s}=\frac{1}{4}\left[(1+\beta_1)\chi_0-\beta_1\alpha_B\right]
\pm\frac{1}{4}\left\{\left[(1-\beta_1)\chi_0+\beta_1\alpha_B\right]^2\right.\nonumber\\
&
&\left.+4\beta_1C_s^2\chi_{\vec{\tau}}\chi_{\vec{-\tau}}\right\}^{-1/2}.
\end{eqnarray}
$\alpha_B=(2\vec{k}\vec{\tau}+\tau^2)k^{-2} $ is the off-Bragg
parameter ($\alpha_B=0$ if the exact Bragg condition of
diffraction is fulfilled),
\[
\gamma_0=\vec{n}_{\gamma}\cdot\vec{N},\quad
\vec{n}_{\gamma}=\frac{\vec{k}}{k},\quad
\beta_1=\frac{\gamma_0}{\gamma_1}, \quad
\gamma_1=\vec{n}_{\gamma\tau}\cdot\vec{N},\quad
\vec{n}_{\gamma\tau}=\frac{\vec{k}+\vec{\tau}}{|\vec{k}+\vec{\tau}|}.
\]
The general solution of (\ref{para_1.15}) inside
a crystal is:
\begin{equation}
\label{para_1.18}
\vec{E}^{(-)s}_{\vec{k}}(\vec{r})=\sum\limits^2_{\mu=1}\left[\vec{e}^s
A_{\mu}\exp(i\vec{k}_{\mu s}\vec{r})+
\vec{e}^s_{\tau}A_{\tau\mu}\exp(i\vec{k}_{\mu
s\tau}\vec{r})\right].
\end{equation}
Associating these solutions with the solutions of Maxwell's
equations for the vacuum area, one can find the explicit form of
$\vec{E}^{(-)s}_{\vec{k}}(\vec{r})$ throughout the space. It is
possible to discriminate several types of diffraction geometries,
namely, the Laue (a) and the Bragg (b) schemes are most well
known.

\textbf{(a)} Let us consider the PXR in the Laue case.

In this case, the electromagnetic waves emitted by a
particle in both the forward and the diffracted directions leave the
crystal through the same surface ($k_z>0, k_z+\tau_z>0$), the
$z$-axis is parallel to the normal $N$ (where $N$ is the normal to
the crystal surface being directed inside a crystal). By matching
the solutions of Maxwell's equations on the crystal surfaces with
the help of (\ref{para_1.15}), (\ref{para_1.17}), (\ref{para_1.18}),
one can obtain the following expressions for the Laue case:
\begin{eqnarray}
\label{para_1.19} & &
\vec{E}^{(-)s}_{\vec{k}}=\left\{\vec{e}^s\left[-\sum_{\mu=1}^2\xi_{\mu
s}^{0*}e^{-i\frac{\omega}{\gamma_0}\varepsilon^*_{\mu
s}L}\right]e^{i\vec{k}\vec{r}}+
e^s_{\vec{\tau}}\beta_1\left[\sum_{\mu=1}^2\xi_{\mu s}^{\tau*}e^{-i\frac{\omega}{\gamma_0}\varepsilon^*_{\mu s}L}\right]e^{i\vec{k}_{\tau}\vec{r}}\right\}\theta(-z)\nonumber\\
& & +\left\{\vec{e}^s\left[-\sum_{\mu=1}^2\xi_{\mu s}^{0*}e^{-i\frac{\omega}{\gamma_0}\varepsilon^*_{\mu s}(L-z)}\right]e^{i\vec{k}\vec{r}}+e^s_{\vec{\tau}}\beta_1\left[\sum_{\mu=1}^2\xi_{\mu s}^{\tau*}e^{-i\frac{\omega}{\gamma_0}\varepsilon^*_{\mu s}(L-z)}\right]e^{i\vec{k}_{\tau}\vec{r}}\right\}\nonumber\\
& & \times \theta(L-z)\theta(z)+\vec{e}^s
e^{i\vec{k}\vec{r}}\theta(z-L),
\end{eqnarray}
where
\begin{eqnarray}
 \xi^0_{1,2 s}=\mp\frac{2\varepsilon_{2,1
s}-\chi_0}{2(\varepsilon_{2s}-\varepsilon_{1s})}; \nonumber
\\ \xi^{\tau}_{1,2
s}=\mp\frac{c_s\chi_{-\tau}}{2(\varepsilon_{2s}-\varepsilon_{1s})};
\nonumber \\
\theta(z)=\left\{ \nonumber
\begin{array}{l}
1, ~~\mbox{if} ~z\geq 0 \\ 0, ~~\mbox{if} ~z<0.
\end{array}
\right.
\end{eqnarray}

Substitution of (\ref{para_1.19}) into (\ref{berk_2.9}) gives for
the Laue case the differential number of quanta of the forward
directed parametric X-rays with the polarization vector
$\vec{e}_s$:
\begin{eqnarray}
\label{para_1.21} & &\frac{d^2N^L_{0s}}{d\omega
d\Omega}=\frac{e^2Q^2\omega}{4\pi^2\hbar c^3}(\vec{e}^s\vec{v})^2
\left|\sum_{\mu=1,2}\xi_{\mu s}^0
e^{i\frac{\omega}{c\gamma_0}\varepsilon_{\mu
s}L}\left[\frac{1}{\omega-\vec{k}\vec{v}}
-\frac{1}{\omega-\vec{k}^*_{\mu s}\vec{v}}\right]\right.\nonumber\\
& &\left.\times[e^{i(\omega-\vec{k}^*_{\mu
s}\vec{v})T}-1]\right|^2,
\end{eqnarray}
where $T=L/c\gamma_0$ is the particle time of flight;
$\vec{e}_1\parallel [\vec{k}\vec{\tau}]$;
$\vec{e}_2\parallel[\vec{k}\vec{e}_1]$.

 One can see that formula (\ref{para_1.21}) looks  like the formula which describes the spectral and angular distribution of the Cherenkov and transition radiations in the matter with the  index of refraction $n_{\mu s}=k_{z\mu s}/k_z=1+\kappa_{\mu s}/k_z$.

The spectral angular distribution for photons in the diffraction
direction $\vec{k}_{\tau}=\vec{k}+\vec{\tau}$ can be obtained from
(\ref{para_1.21}) by a simple substitution
\begin{eqnarray*}
& &\vec{e}_s\rightarrow\vec{e}_{s\tau},\qquad \xi^0_{\mu s}\rightarrow \beta_1\xi_{\mu s}^{\tau},\\
& &\xi^{\tau}_{1(2)s}=\pm\frac{\chi_{\tau}c_s}{2(\varepsilon_{1s}-\varepsilon_{2s})}\\
& &\vec{k}\rightarrow\vec{k}_{\tau}, \quad\vec{k}_{\mu
s}\rightarrow\vec{k}_{\tau\mu s}=\vec{k}_{\mu s}+{\tau}.
\end{eqnarray*}

\textbf{(b)} Now let us consider PXR in the Bragg case. In this
case, side by side with the electromagnetic wave emitted in the
forward direction, the electromagnetic wave emitted by a charged
particle in the diffracted direction and leaving the crystal
through the surface of the particle entrance can be observed. By
matching the solutions of Maxwell's equations on the crystal
surface with the help of (\ref{para_1.15}), (\ref{para_1.17}),
(\ref{para_1.18}), one can get the formulas for the Bragg
diffraction schemes.

It is interesting that the spectral angular distribution for
photons emitted in the forward direction can be obtained from
(\ref{para_1.21}) by the following substitution, $\xi^0_{\mu
s}\rightarrow \gamma_{\mu s}$,
\begin{equation}
\label{para_1.22} \gamma^0_{1(2)s}=
 \frac{2\varepsilon_{2(1)s}-\chi_0}{(2\varepsilon_{2(1)s}-\chi_0)
-(2\varepsilon_{1(2)s}-\chi_0)
e^{i\frac{\omega}{\gamma_0}(\varepsilon_{2(1)s}-\varepsilon_{1(2)s})L}}
\end{equation}

The spectral angular distribution of photons emitted in the
diffracted direction can be obtained from (\ref{para_1.21}) by
substitution
\begin{eqnarray*}
\vec{e}_s\rightarrow\vec{e}_{s\tau},\quad
\vec{k}\rightarrow\vec{k}_{\tau}, \quad k_{\mu
s}\rightarrow\vec{k}_{\mu\tau s}, \quad
\xi^0_{\mu s}e^{i\frac{\omega}{\gamma_0}\varepsilon_{\mu
s}L}\rightarrow \gamma^{\tau}_{\mu s},
\end{eqnarray*}
where
\begin{equation}
\gamma^{\tau}_{1(2)s}= - \frac{\beta_1 c_s\chi_{\tau}}
{(2\varepsilon_{2(1)s}-\chi_0)-(2\varepsilon_{1(2)s}-\chi_0)e^{i\frac{\omega}{\gamma_0}(\varepsilon_{2(1)s}-\varepsilon_{1(2)s})L}}.
\end{equation}

Let us note that the above formulas fully describe parametric (quasi-Cherenkov) radiation in natural and photonic crystals and they certainly include that contribution to radiation, which goes over to ordinary transition radiation, if the radiation is considered outside the region of diffraction reflection. A description of diffracted radiation of a relativistic oscillator is given in \cite{1,2} and the reference therein.

Let us take notice of the fact that in photonic crystals built
from metal threads with the diameter smaller than or comparable
with $\lambda$, the value of $\chi(\tau)$ is practically
independent on $\tau$. As a result, it is possible to effectively
excite radiation in, e.g.,  the terahetrz range in a lattice with
a period of several millimeters.

When a particle travels in a vacuum near the surface of a
spatially periodic medium, new kinds of radiation arise
\cite{berk_104,berk_105} -- surface parametric (quasi-Cherenkov)
X-ray radiation (SPXR) and surface DRO (see Figure \ref{berkley
Figure 30}). This phenomenon takes place under the condition of
uncoplanar  surface diffraction, first considered in \cite{147}.

\begin{figure}[htp]
\centering
\epsfxsize = 8 cm \centerline{\epsfbox{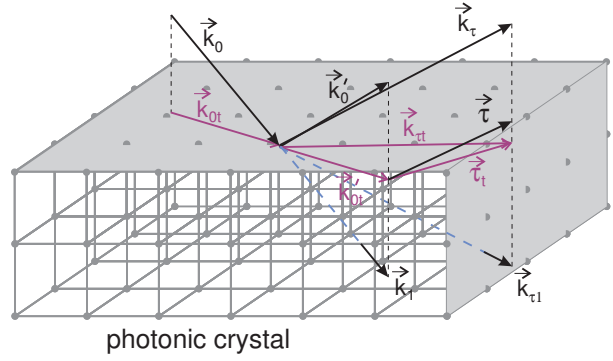}}
\caption{Surface diffraction of a radiated photon} \label{berkley
Figure 30}
\end{figure}

The solution of Maxwell's equation $\vec{E}_{\vec{k}}^{(+)}(\vec{r})$ in this case of uncoplanar  surface diffraction was obtained in \cite{147}. It was shown that the surface diffraction in the two-wave case is characterized by two angles of total reflection (several angles in the case  of multi-wave diffraction \cite{berk_107}). The solution obtained in \cite{berk_107} contains the component, which describes the state that damps with growing distance from the surface of the medium, both within the material and in the vacuum, and which describes a surface wave, i.e., a wave in which the energy flux is directed along the boundary of the surface of a spatially periodic target (see review \cite{berk_108}). According to \cite{147}, this solution, which describes scattering of a plane wave by the target under the surface diffraction geometry, can be written in the form:
\begin{equation}
\label{berk_2.63}
\vec{E}^{(+)s}_{\vec{k}}=e_s e^{i\vec{k}\vec{r}}+A_s(\vec{k}, \omega)e^{i\vec{k}_1\vec{r}}+B_s(\vec{k},\omega)e^{i\vec{k}_2\vec{r}},
\end{equation}
where the wave vector in a vacuum $\vec{k}=(\vec{k}_t,
\vec{k}_{\perp})$, $\vec{k}_1=(\vec{k}_t,- \vec{k}_{\perp})$,
$\vec{k}_2=(\vec{k}_{2t},- \vec{k}_{2\perp})$,
$|\vec{k}_{2\perp}|=\sqrt{k^2-k^2_{2t}}$, $\vec{k}_{2t}=\vec{k}_t+
\vec{\tau}$, $\vec{k}_t$ is the component of the wave vector that
is parallel to the surface, $\vec{\tau}$ is the reciprocal lattice
vector, $\omega$ is the photon frequency. The amplitudes $A_s$ and
$B_s$ are  given in \cite{7,berk_105}. Substituting the solution
$\vec{E}^{(-)s}_{\vec{k}}=(\vec{E}^{(+)s}_{-\vec{k}})^*$ into
(\ref{berk_2.3}), we can find the spectral-angular distribution of
SPXR and DRO.


\section{Time dependence of the intensity of radiation produced by a particle transmitted through a crystal}

The intensity $I(t)$ of radiation produced by a particle which has passed through a crystal can be found with known intensity of the electric field $\vec{E}(\vec{r},t))$  (magnetic field $\vec{H}(\vec{r},t)$) of the electromagnetic wave, which is produced by this particle \cite{L_14},
\begin{equation}
\label{1.1}
I(t)=\frac{c}{4\pi}|\vec{E}(\vec{r},t)|^2r^2d\Omega,
\end{equation}
where $r$ is the distance from the crystal, which is assumed to be larger than the crystal size.

The field $\vec{E}(\vec{r},t)$ can be presented as an expansion in a Fourier series
\begin{equation}
\label{1.2}
\vec{E}(\vec{r},t)=\frac{1}{2\pi}\int\vec{E}(\vec{r},\omega)e^{-i\omega t}d\omega.
\end{equation}
According to the results obtained in \cite{3,5,7},  at a long distance from the crystal, the Fourier component can be written as follows:
\begin{equation}
\label{1.3}
\vec{E}(\vec{r},t)=\frac{e^{ikr}}{r}\frac{i\omega}{c^2}\sum_{s}e_i^s
\int\vec{E}_{\vec{k}}^{(-)s^*}(\vec{r}^{\prime}\omega)\vec{j}(\vec{r}^{\prime},\omega)  d^3 r^{\prime}.
\end{equation}
where $i=1,2,3$ (and correspond(s) to the coordinate axes $x$, $y$, $z$), $e^s_i$ is the $i$-component  of the wave polarization vector $\vec{e}^s$; $s=1,2$; $\vec{E}^{(-)s}_{\vec{k}}$ is the solution of  Maxwell's equations describing scattering of a plane wave with a wave vector $\vec{k}=k\frac{\vec{r}}{r}$ and the asymptotic of a converging spherical wave,
\begin{equation}
\label{1.4}
\vec{j}(\vec{r}, \omega)=\int \vec{j}(\vec{r}, t)e^{i\omega t}dt
\end{equation}
$\vec{j}(\vec{r}, \omega)= Q\vec{v}(t)\delta(\vec{r}-\vec{r}(t))$ is the is the current density of the particle with charge $Q$, $\vec{r}(t)$ is the particle coordinate at time $t$.

The explicit form of the expressions $\vec{E}^{(-)s}$ describing
diffraction of the electromagnetic wave in a crystal in the Laue
and Bragg cases is given in \cite{3,6,lanl_7a} (See Section 1).

Now let us take a closer look at the expression for the amplitude $A(\omega)$ of the emitted wave:
\begin{equation}
\label{1.5}
A_{\vec{k}}^s(\omega)=\frac{i\omega}{c^2}\int \vec{E}_{\vec{k}}^{(-)s^*}(\vec{r}^{\prime},\omega)\vec{j}(\vec{r}^{\prime},\omega)d^3 r^{\prime}.
\end{equation}
Using (\ref{1.4}), (\ref{1.5}) can be recast as follows
\begin{eqnarray}
\label{1.61}
A_{\vec{k}}^s(\omega)&=&\frac{i\omega}{c^2}\int \vec{E}_{\vec{k}}^{(-)s^*}(\vec{r}^{\prime},\omega)Q\vec{v}(t)\delta(\vec{r}^{\prime}-\vec{r}(t))e^{i\omega t}dt d^3 r^{\prime}\nonumber\\
&=&\frac{i\omega Q}{c^2}\int\vec{E}_{\vec{k}}^{(-)s^*}(\vec{r}(t),\omega)\vec{v}(t)e^{i\omega t} dt
\end{eqnarray}

Recall that
$\vec{E}_{\vec{k}}^{(-)s^*}=\vec{E}_{-\vec{k}}^{(+)s}$, where the
field $\vec{E}_{-\vec{k}}^{(+)s}$ is the solution of Maxwell's
equations describing scattering by a crystal of a plane wave with
wave vector $(-\vec{k})$ and the asymptotics of a diverging wave
at infinity. According to (\ref{1.61}), the radiation amplitude is
determined by the field $\vec{E}_{\vec{k}}^{(-)s}$ taken at point
$\vec{r}(t)$ of particle location at time $t$ and integrated over
the time of particle motion.

Let us  consider in more detail the constant motion of a particle
in passing through the crystal. In this case,  parametric
quasi-Cherenkov radiation can appear \cite{1,6}, which includes,
as a particular case, diffracted transition radiation. The
explicit formulas for the radiation amplitude in the case of
two-wave diffraction of photons in crystals for the Laue and Bragg
geometries are given in \cite{3,6,lanl_7a} (Section 1).

From (\ref{1.2}), (\ref{1.3}), and (\ref{1.5}) follows that the expression for the electromagnetic wave emitted by the particle passing through the crystal (natural or photonic) can be presented in a form:
\begin{equation}
\label{1.7}
\vec{E}_i(\vec{r},t)=\frac{1}{2\pi r}\sum_s e^s_i\int A_{\vec{k}}^s(\omega)e^{-i\omega(t-\frac{r}{c})}d\omega,
\end{equation}
i.e., $\vec{E}_i(\vec{r},t)=\frac{1}{r}\sum\limits_s e_i^s A^s_{\vec{k}}(t-\frac{r}{c})$.

From (\ref{1.7}) follows that the time dependence of the form of the pulse $I(\vec{r},t)(\vec{E}(\vec{r},t))$ of radiation generated by a particle passing through the crystal is determined by the dependence of the radiation amplitude $A_{\vec{k}}^s(\omega)$ on frequency. According to the explicit expression for the radiation amplitudes given in  \cite{3,6,lanl_7a}, the radiation amplitudes  $A_{\vec{k}}^s(\omega)$ can be presented as sums proportional to the amplitudes of diffraction reflection from the crystal and to the amplitude of wave transmission through the crystal. For example, for the case of forward parametric radiation  in the Laue geometry
\begin{eqnarray}
\label{1.6}
A_{\vec{k}}^s(\omega)&=&\frac{Q}{c^2}(\vec{e}^s \vec{v})\sum_{\mu=1,2}\xi^0_{\mu s}e^{i\frac{\omega}{\gamma_0}\varepsilon_{\mu s}L}\nonumber\\
&\times&\left[\frac{1}{\omega-\vec{k}\vec{v}}-\frac{1}{\omega-(\vec{k}+\kappa_{\mu s}\vec{N})\vec{v}}\right]\left[e^{i(\omega-(\vec{k}+\kappa_{\mu s}\vec{N})\vec{u})\frac{L}{c\gamma_0}} -1\right]
\end{eqnarray}
Thus, the time dependence of the from of the radiation pulse is determined by the time dependence of the radiation amplitude $A_{\vec{k}}^s(t-\frac{r}{c})$.

By way of example, let us consider the characteristics of the time
dependence of radiation produced by a particle passing through the
crystal for a wave packet passing through the crystal
\cite{2,nan,nan_lanl}


Let us consider the pulse of electromagnetic radiation passing through the
medium with the index of refraction $n(\omega )$. The  group
velocity of the wave packet is as follows:

\begin{equation}
v_{gr}=\left( \frac{\partial \omega n(\omega )}{c\partial \omega }\right)
^{-1}=\frac{c}{n(\omega )+\omega \frac{\partial n(\omega )}{\partial \omega }%
},
\label{v1}
\end{equation}
where $c$  is the speed of light, $\omega $  is the quantum frequency.

In the X-ray range ( $\sim $tens of keV) the index of refraction has the
universal form $n(\omega )=1-\frac{\omega _{L}^{2}}{2\omega ^{2}}$ , $\
\omega _{L}$ is the Langmuir frequency. Additionally, $n-1\simeq 10^{-6}\ll 1
$. Substituting \ $n(\omega )$ into (\ref{v1}), one can obtain that $v_{gr}\simeq
c\left( 1-\frac{\omega _{L}^{2}}{\omega ^{2}}\right) $. It is clear that
the group velocity is close to the speed of light. Therefore the time delay of
the wave packet  in a medium is much shorter than the time needed for
passing the path equal to the target thickness in a vacuum.

\begin{equation}
\Delta T=\frac{l}{v_{gr}}-\frac{l}{c}\simeq \frac{l}{c}\frac{\omega _{L}^{2}%
}{\omega ^{2}}\ll \frac{l}{c}.
\label{v2}
\end{equation}

To consider the pulse diffraction in a crystal, one should solve Maxwell's
equations that describe a pulse passing through a crystal. Maxwell's equations
are linear, therefore it is convenient to use the Fourier transform in time and
to rewrite these equations as functions of frequency:
\begin{equation}
\left[ -curl~curl~\vec{E}_{\vec{k}}(\vec{r},\omega )+\frac{\omega ^{2}}{c^{2}%
}\vec{E}_{\vec{k}}(\vec{r},\omega )\right] _{i}+\chi _{ij}(\vec{r},\omega
)~E_{\vec{k},j}(\vec{r},\omega )=0,
\label{v3}
\end{equation}
where $\chi _{ij}(\vec{r},\omega )$  is the spatially periodic
tensor of susceptibility; $i,j=1,2,3$ repeated  indices imply
summation.

Making the Fourier transformation of these equations in coordinate
variables, one can derive a set of equations associating the
incident and diffracted waves. When two strong waves are excited
under diffraction (the so-called two-beam diffraction case), the
following set of equations for determining the wave amplitudes can
be obtained:
\begin{equation}
\begin{array}{c}
\left( \frac{k^{2}}{\omega ^{2}}-1-\chi _{0}\right) \vec{E}_{\vec{k}%
}^{s}-c_{s}\chi _{-\vec{\tau}}\vec{E}_{\vec{k}_{\tau }}^{s}=0 \\
\\
\left( \frac{k_{\tau }^{2}}{\omega ^{2}}-1-\chi _{0}\right) \vec{E}_{\vec{k}%
_{\tau }}^{s}-c_{s}\chi _{\vec{\tau}}\vec{E}_{\vec{k}}^{s}=0
\label{v4}
\end{array}
\end{equation}
Here $\vec{k}$ is the wave vector of the incident wave, $\vec{k}_{\vec{\tau}%
}=\vec{k}+\vec{\tau}$, $\vec{\tau}$ is the reciprocal lattice vector; $\chi
_{0},\chi _{\vec{\tau}}$ are the Fourier components of the crystal
susceptibility:
\begin{equation}
\chi (\vec{r})=\sum_{\vec{\tau}}\chi _{\vec{\tau}}\exp (i\vec{\tau}\vec{r})
\label{v5}
\end{equation}
$C_{s}=\vec{e}^{~s}\vec{e}_{\vec{\tau}}^{~s}$, $\vec{e}^{~s}(\vec{e}_{\vec{%
\tau}}^{~s})$ are the unit polarization vectors of the incident and
diffracted waves, respectively.

The solvability condition for the linear system (\ref{v4}) leads to a dispersion
equation that determines the possible wave vectors $\vec{k}$ in a crystal.
It is convenient to present these wave vectors as:
\[
\vec{k}_{\mu s}=\vec{k}+\texttt{\ae }_{\mu s}\vec{N},~\ae _{\mu s}=\frac{%
\omega }{c\gamma _{0}}~\varepsilon _{\mu s},
\]
where $\mu =1,2$; $\vec{N}$ is the unit vector of a normal to the entrance surface
of the crystal, which is directed into the crystal,
\begin{equation}
\varepsilon _{s}^{(1,2)}=\frac{1}{4}[(1+\beta )\chi _{0}-\beta \alpha _{B}%
]\pm \frac{1}{4}\left\{ [(1+\beta )\chi _{0}-\beta \alpha _{B}-2\chi
_{0}]^{2}+4\beta C_{S}^{2}\chi _{\vec{\tau}}\chi _{-\vec{\tau}}\right\}
^{1/2},
\label{v6}
\end{equation}
$\alpha _{B}=(2\vec{k}\vec{\tau}+\tau ^{2})k^{-2}$ is the off-Bragg
parameter ($\alpha _{B}=0$ when the Bragg condition of diffraction is
exactly fulfilled),
\[
\gamma _{0}=\vec{n}_{\gamma }\cdot \vec{N},~~~\vec{n}_{\gamma }=\frac{\vec{k}%
}{k},~~~\beta =\frac{\gamma _{0}}{\gamma _{1}},~~~\gamma _{1}=\vec{n}%
_{\gamma \tau }\cdot \vec{N},~~~\vec{n}_{\gamma \tau }=\frac{\vec{k}+\vec{%
\tau}}{|\vec{k}+\vec{\tau}|}
\]
The general solution of equations (\ref{v3}), (\ref{v4}) inside a crystal is:
\begin{equation}
\vec{E}_{\vec{k}}^{s}(\vec{r})=\sum_{\mu =1}^{2}\left[ \vec{e}^{~s}A_{\mu
}\exp (i\vec{k}_{\mu s}\vec{r})+\vec{e}_{\tau }^{~s}A_{\tau \mu }\exp (i\vec{%
k}_{\mu s\tau }\vec{r})\right]
\label{v7}
\end{equation}

Associating these solutions with the solutions of Maxwell's
equation for the
vacuum area one can find the explicit expression for $\vec{E}_{\vec{k}}^{s}(%
\vec{r})$ throughout the space. It is possible to discriminate several types
of diffraction geometries, namely, the Laue and the Bragg schemes, which  are
most well-known \cite{nan4}.

In the case of two-wave dynamical diffraction, the crystal can be described \ by
two effective  indices of refraction

\[
n_{s}^{(1,2)}=1+\varepsilon _{s}^{(1,2)},
\]

\begin{equation}
\varepsilon _{s}^{(1,2)}=\frac{1}{4}\left\{ \chi _{{\small 0}}(1+\beta
)-\beta \alpha \pm \sqrt{(\chi _{{\small 0}}(1-\beta )+\beta \alpha
)^{2}+4\beta C_{s}\chi _{{\small \tau }}\chi _{{\small -\tau }}}\right\} .
\label{v8}
\end{equation}

The diffraction is significant in the narrow range near the Bragg frequency,
therefore $\chi _{0}$ and $\chi _{\tau }$ can be considered as constants
and\ the dependence on $\omega $ should be taken into account for \ $\alpha =%
\frac{2\pi \overrightarrow{\tau }(2\pi \overrightarrow{\tau }+2%
\overrightarrow{k})}{k^{2}}=-\frac{(2\pi \tau )^{2}}{k_{B}^{3}c}(\omega
-\omega _{B})$, where $k=\frac{\omega }{c}$; $2\pi \overrightarrow{\tau }$\
is the reciprocal lattice vector which characterizes the set of planes where
the diffraction occurs; Bragg frequency is determined by the condition $%
\alpha =0$.

From (\ref{v1}), (\ref{v8})\ one can obtain

\begin{equation}
v_{gr}^{(1,2)s}=\frac{c}{n^{(1,2)}(\omega )\pm \beta \frac{(2\pi \tau )^{2}%
}{4k_{B}^{2}}\frac{(\chi _{{\small 0}}(1-\beta )+\beta \alpha )}{\sqrt{%
(\chi _{0}(1-\beta )+\beta \alpha )^{2}+4\beta C_{s}\chi _{
\tau }\chi _{ -\tau }}}}.
\label{v9}
\end{equation}

\bigskip In the general case $(\chi _{0}(1-\beta )+\beta \alpha )\simeq 2%
\sqrt{\beta }\chi _{0}$, therefore the term that is added to $%
n_{s}^{(1,2)}(\omega )$ in the denominator (\ref{v9}) is of the order of 1.\
Moreover, $v_{gr}$ significantly differs from $c$ for the antisymmetric
diffraction $(\left| \beta \right| \gg 1).$ It should be noted that because
of the complicated character of the wave field in a crystal, one of  $%
v_{gr}^{(i)s}$ can appear to be much higher than $c$\ and negative. When $%
\beta $ is negative the radicand in (\ref{v9}) can become zero
(Bragg reflection threshold) and $v_{gr}\rightarrow 0$\ . It
should be noted that in the presence of a variable external field,
a crystal can be described by the effective indices of refraction
which depend on the external field frequency $\Omega $\ .
Therefore in this case $v_{gr}$ appears to be the function of
$\Omega $\ . This can be easily observed in the conditions of
X-ray-acoustic resonance. The performed analysis allows one to
conclude that the center of the X-ray pulse in a crystal can \
undergo a significant delay $\Delta T\gg \frac{l}{c}$ available
for experimental investigation. Thus, when $\beta =10^{3}$,
$l=0.1$ cm and $l/c\simeq 3\cdot 10^{-12}$, the delay \ time can
be estimated as $\Delta T\simeq 3\cdot 10^{-9} $sec.

\bigskip

Let us study now the time dependence of the delay law of radiation \ after
passing through a crystal. Assuming that $B(\omega )$ is the reflection or
transmission amplitude coefficients of a crystal, one can obtain the
following expression for the pulse form

\begin{equation}
E(t)=\frac{1}{2\pi }\int B(\omega )E_{0}(\omega )e^{-i\omega t}d\omega =\int
B(t-t^{\prime })E_{0}(t^{\prime })dt^{\prime }.
\label{v10}
\end{equation}
where $E_{0}(\omega )$ is the amplitude of the electromagnetic wave incident
on a crystal\ \ \ \

\bigskip In accordance with the general theory, for the Bragg geometry, the
amplitude of the diffraction-reflected wave for the crystal width
 much greater than the absorbtion length can be written as
 \cite{nan4}:

\begin{eqnarray}
& &B_{s}(\omega )=\\
& &-\frac{1}{2\chi _{\tau }}\left\{ \chi _{{\small 0}}(1+\left|
\beta \right| )-\left| \beta \right| \alpha -\sqrt{(\chi _{{\small 0}%
}(1-\left| \beta \right| )-\left| \beta \right| \alpha )^{2}-4\left| \beta
\right| C_{s}\chi _{{\small \tau }}\chi _{{\small -\tau }}}\right\}\nonumber
\label{v11}
\end{eqnarray}

\bigskip
In the absence of resonance scattering, the parameters $\chi _{0}$
and $\chi _{\pm \tau }$ can be considered as \ constants and frequency
dependence is defined by the term $\alpha =-\frac{(2\pi \tau )^{2}}{%
k_{B}^{3}c}(\omega -\omega _{B})$. So, $B_{s}(t)$\ \ can be found from

\begin{eqnarray}
& &B_{s}(t)=-\frac{1}{4\pi \chi _{\tau }}\\
& &\times \int \left\{ \chi _{{\small 0}%
}(1+\left| \beta \right| )-\left| \beta \right| \alpha -\sqrt{(\chi _{%
{\small 0}}(1-\left| \beta \right| )-\left| \beta \right| \alpha
)^{2}-4\left| \beta \right| C_{s}\chi _{{\small \tau }}\chi _{{\small -\tau }%
}}\right\} e^{-i\omega t}d\omega .\nonumber
\label{v12}
\end{eqnarray}

The Fourier transform of the first term results in $\delta (t)$ and we can
neglect it because the delay is described by the second term. The second
term can be calculated by the methods of the theory of \ function of complex
argument:

\begin{equation}
B_{s}(t)=-\frac{i}{4\chi _{\tau }}\left| \beta \right| \frac{(2\pi \tau
)^{2}}{k_{B}^{2}\omega _{B}}\frac{J_{1}(a_{s}t)}{t}e^{-i(\omega _{B}+\Delta
\omega _{B})t}\theta (t),
\label{v13}
\end{equation}

\bigskip or

\begin{equation}
B_{s}(t)=-\frac{i\sqrt{\left| \beta \right| }}{2}\frac{J_{1}(a_{s}t)}{a_{s}t}%
e^{-i(\omega _{B}+\Delta \omega _{B})t}\theta (t),
\label{v14}
\end{equation}

where

\[
a_{s}=\frac{2\sqrt{C_{s}\chi _{\tau }\chi _{-\tau }}\omega _{B}}{\sqrt{%
\left| \beta \right| }\frac{(2\pi \tau )^{2}}{k_{B}^{2}}},\Delta \omega
_{B}=-\frac{\chi _{{\small 0}}(1+\left| \beta \right| )\omega _{B}k_{B}^{2}}{%
\left| \beta \right| (2\pi \tau )^{2}}.
\]

\bigskip

Since $\chi _{0}$ and $\chi _{\tau }$ are complex, both $a_{s}$
and $\Delta \omega _{B}$ have real and imaginary parts. According
to (\ref{v12})--(\ref{v14}), in the case of Bragg reflection of a
short pulse (the pulse frequency bandwidth $\gg $ frequency
bandwidth of the total reflection range)  both the instantly
reflected pulse and the pulse with amplitude undergoing damped
beatings appear. Beatings period increases with $\left| \beta
\right| $ grows and $\chi _{\tau }$\ decrease. Pulse intensity can
be written as

\begin{equation}
I_{s}(t)\sim \left| B_{s}(t)\right| ^{2}=\frac{\left| \beta \right| }{2}%
\left| \frac{J_{1}(a_{s}t)}{at}\right| ^{2}e^{-2\texttt{Im}\Delta \omega
_{B}t}\theta (t).
\label{v15}
\end{equation}

It is evident that the reflected pulse intensity depends on the orientation
of photon polarization vector $\vec{e}_{s}$ and undergoes the
damping oscillations on time. \qquad \qquad\ \

Let us evaluate the effect. Characteristic values are $\texttt{Im}\Delta
\omega _{B}\sim \texttt{Im}\chi _{0}\omega _{B}$ and $\texttt{Im}a\sim \frac{%
\texttt{Im}\chi _{\tau }\omega _{B}}{\sqrt{\beta }}.$ For 10 keV for the
crystal of Si $\ \texttt{Im}\chi _{0}=1,6\cdot 10^{-7}$ , \ for LiH $\ \texttt{Im%
}\chi _{0}=7,6\cdot 10^{-11},\texttt{Im}\chi _{\tau }=7\cdot 10^{-11}$, \ for
LiF $\ \texttt{Im}\chi _{0}\sim 10^{-8}.$ Consequently, the characteristic
time $\tau $\ for the exponent decay in (\ref{v15}) can be estimated as follows ($%
\omega _{B}=10^{19}$):

for Si the characteristic time $\tau \sim 10^{-12}$ sec, for LiF
the characteristic time $\tau \sim 10^{-10}$ sec, for LiH the
characteristic time $\tau \sim 10^{-9}$ sec!!

The reflected pulse also undergoes oscillations, the period of
which increases with growing $\left| \beta \right| $ and
decreasing $\texttt{Re}\chi _{\tau }. $ This period can be
estimated for $\beta =10^{2}$ and $\texttt{Re}\chi _{\tau }\sim
10^{-6}$ as $T \sim 10^{-12}$ sec (for Si, LiH, LiF).

When the resolving time of the detecting equipment is greater than
the oscillation period, the expression (\ref{v15}) should be
averaged over the
period of oscillations. Then, for the time intervals when $\texttt{Re}%
a_{s}t\gg 1,$ $\texttt{Im}\Delta \omega _{B}t\ll 1$ the delay law (\ref{v15}) has the
power function form:

\[
{\large I}_{s}{\large (t)\,\sim \,t}^{-3}{\large .}
\]


 In  the case of multi-wave diffraction, the time delay for the photon exit from the crystal will be even more appreciable.

 For an artificial spatially periodic medium (diffraction grating, photonic crystal), the parameter $g_{0\,1}$ can vary over a wide range.
 For example, according to \cite{nim06}, for a photonic crystal built from tungsten threads of 100$\mu m$ in diameter,
 the parameter $g_{0\,1}\sim\frac{1}{\omega^2}$ has the value of $g_{0\,1}\sim 10^{-2}$ in a 10 GHz range.
 As a result,
 in this range we have $T$ (10 GHz)$\sim\frac{\sqrt{\beta}}{|g_{0\,1}|\omega_B}\sim\sqrt{\beta}\cdot 10^{-9}$.
 At the same time, in the terahertz range, $T$ (1 THz) due to the drop of $g_{0\,1}$ ($T$ increases
 proportionally to $\omega$, the parameter $a$ decreases: $a\sim\frac{1}{\omega_B}$ ), we have
 the period $T$ (1 THz) $\sim\sqrt{\beta}\cdot 10^{-6}$. As is seen, the oscillations of radiation
 from photonic crystals are quite observable.

\begin{figure}[htp]
\centering \epsfxsize = 8 cm \centerline{\epsfbox{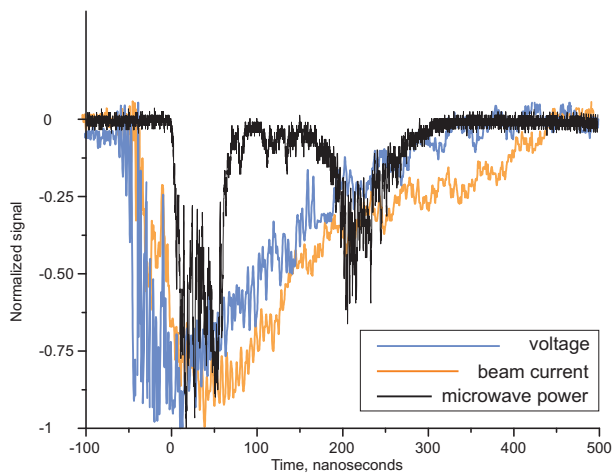}}
\caption{Detected microwave signal (black curve) synchronized with
the beam current and electron gun voltage} \label{fig:exp}
\end{figure}

 So the time $\tau_{ph}=\frac{L}{v_{gr}}$ that the photon spends in the crystal can be longer than the flight time $\tau_p=\frac{L}{v}$ of a
 relativistic particle in a crystal. Hence, the emission of diffraction-related radiation (quasi-Cherenkov,
 transition, diffracted radiation of an oscillator, surface parametric radiation and others) produced by a
 relativistic particle will continue after the particle has left the crystal (see
 Fig.\ref{fig:exp})
 Under diffraction conditions,
 the crystal acts as a high-quality resonator \cite{1,Q-factor}.

 It should be noted, of course, that  in observation of oscillations, one should either register the moment of particle entrance into the crystal or use a short bunch of
  particles with duration much shorter than the oscillation period. In the X-ray range, such situation is
  typical of electron buches, which are applied for creating X-ray FELs (DESY). (The bunch duration in such
  FELs is tens-hundreds of femptoseconds). In the terahertz range, much longer bunches are required, so there
  are not serious experimental problems in this case. If the bunch duration is large in comparison with the duration
  of the radiation pulse or the time of the electron entrance into the crystal is not registered, which occurs in a conventional
  experimental arrangement, then the intensity $I(t)$  should be integrated over
  longer  observation time intervals. As a result, we, in fact, obtain the expression (\ref{berk_2.1})
  integrated over all frequencies, i.e., an ordinary stationary angular distribution of radiation.
  If the response time  of the devices detecting $\tau_D$ (or  the flight time of the particle in a crystal,
  or the bunch duration) is comparable with the oscillation period, then $I(t)$ should be integrated over the interval
  $\tau_D$. In this case oscillations will disappear, but we will observe the power-law decrease in the intensity of radiation from the crystal.

In according with the above analysis some experiments are carried
to observe delay of radiation pulse in a photonic crystal used for
VFEL lasing \cite{FEL06,FEL09,FEL10,IRMMW10}. In these experiments
the additional radiation peak (see
 Fig.\ref{fig:exp}) is observed at studies of lasing
of VFEL with ''grid'' photonic crystals in backward wave
oscillator regime. This peak appears when the electron beam has
left the resonator.

It should be mentioned here that backward wave oscillator regime
implies generation in presence of Bragg diffraction, therefore,
under some conditions the group velocity could  appear even to be
close to 0 (see equation (\ref{v9})). The observed delay
(Fig.\ref{fig:exp}) corresponds to $v_{gr} \sim 10^8$ cm/s, i.e.
$\frac{v_{gr}}{c} \sim 10^{-2}$.

In travelling wave regime, which corresponds to case of Laue
diffraction, such long delay can not be obtained (according to
(\ref{v9}) for $\beta > 0$ the group velocity $v_{gr}$ changes
insignificantly). Particularly, in our experiments with Cherenkov
generator without diffraction grating no additional peaks are
detected, because the group velocity in this case changes
insignificantly due to the same reasons as in the Laue case.

And after all note that diffraction of a pulse of radiation
produced by an external radiation source in a periodic structure
could be accompanied by appearance of several transmitted or
reflected radiation pulses (pulses of photons) (see
\cite{VG+Maks}).

\section*{Conclusion}
The formulas which describe the time evolution of radiation
produced by a relativistic particle moving in a crystal are
derived. It is shown that the conditions are realizable under
which parametric (quasi-Cherenkov) radiation, transition
radiation, diffracted radiation of the oscillator, surface
quasi-Cherenkov and Smith-Purcell radiation last considerably
longer than the time $\tau_p$ of  the particle flight through the
crystal. The results of carried out experiments  demonstrate the
presence of additional radiation peak appearing after the electron
beam has left the photonic crystal.

\end{document}